# A fundamental numerical and theoretical study for the vibrational properties of nanowires

H.F. Zhan and Y.T. Gu*

*School of Chemistry, Physics and Mechanical Engineering, Queensland University of Technology,*

*Brisbane, 4001, Australia*

**\*Corresponding Author:** Dr. Yuantong Gu

**Mailing Address:** School of Chemistry, Physics and Mechanical Engineering,

Queensland University of Technology,

GPO Box 2434, Brisbane, QLD 4001, Australia

**Telephones:** +61-7-31381009

**Fax:** +61-7-31381469

**E-mail:** yuantong.gu@qut.edu.au

**Abstract:** Based on the molecular dynamics (MD) simulation and the classical Euler-Bernoulli beam theory, a fundamental study of the vibrational performance of the Ag nanowire (NW) is carried out. A comprehensive analysis of the quality (Q)-factor, natural frequency, beat vibration, as well as high vibration mode is presented. Two excitation approaches, i.e., velocity excitation and displacement excitation, have been successfully implemented to achieve the vibration of NWs. Upon these two kinds of excitations, consistent results are obtained, i.e., the increase of the initial excitation amplitude will lead to a decrease to the Q-factor, and moderate plastic deformation could increase the first natural frequency. Meanwhile, the beat vibration driven by a single relative large excitation or two uniform excitations in both two lateral directions is observed. It is concluded that the nonlinear changing trend of external energy magnitude does not necessary mean a non-constant Q-factor. In particular, the first order natural frequency of the Ag NW is observed to decrease with the increase of temperature. Furthermore, comparing with the predictions by Euler-Bernoulli beam theory, the MD simulation provides a larger and smaller first vibration frequency for the clamped-clamped and clamped-free thin Ag NWs, respectively. Additionally, for thin NWs, the first order natural frequency exhibits a parabolic relationship with the excitation magnitudes. The frequencies of the higher vibration modes tend to be low in comparison to Euler-Bernoulli beam theory predictions. A combined initial excitation is proposed which is capable to drive the NW under a multi-modes vibration and arrows the coexistence of all the following low vibration modes. This work sheds lights on the better understanding of the mechanical properties of NWs, and benefits the increasing utilities of NWs in diverse nano-electronic devices.

## I. INTRODUCTION

Driven by their remarkable mechanical, electrical, optical, and other properties, nanowires (NWs) have drawn considerable interests from the scientific community. Their intriguing properties have enabled them being widely applied as building blocks of nanoelectromechanical systems (NEMS), such as high frequency resonator,[1, 2] force and pressure sensing,[3] field effect transistors (FETs),[4] and other devices.[5-8] These nanowire-based devices are expected to quickly find their way into high-performance electronics and commercial products, and could improve the performance of mobile phones, radars and other wireless systems.[9] Among the NEMS, NWs are commonly utilized as a resonating beam, in which they are often set to oscillate continuously at or near its resonant frequency, thus, the changes in local environment, such as force, pressure or mass can be detected. This schematic has been popularly applied in atomic force microscopy (AFM) and various kinds of sensors and actuators.[10] Therefore, it is of great scientific interests to characterize the NW's mechanical properties under vibration.

For the experimental vibration study, an ac electric-filed excitation is typically adopted to drive the mechanical vibration of a NW at its first resonance frequency, which is often carried out for a cantilevered NW residing in a sample. This technique has been successfully implemented by Dikin et al.,[11] who studied the bending modulus and electron charge trapping effect on amorphous $SiO_2$ NWs. One of the key performance measurements during vibration is the quality (Q)-factor, which refers the energy dissipation rate during each vibrational circle. It is confirmed that, higher Q-factor means higher sensitivity, and more reliable performance of the NEMS.[12] Another crucial vibrational property is the resonance frequency, which could be deduced from the governing differential equation according to the beam theory.

In the past few years, several studies of the Q-factor of different kinds of NWs have been conducted. For instance, a high Q-factor was reported for GaN NW cantilevers grown from molecular beam epitaxy,[13] as well as batch-fabricated SiC NWs.[14] A recent experimental study reveals that the wavelength and relative intensity of the resonance models in the CdS NW cavity could be tuned by adjusting the relative position of the Ag nanoparticle.[15] Some theoretical works have also being carried out considering the effects of surface stress on the natural frequency of NWs, e.g., the incorporation of surface effect in the classical beam theories.[16-18] Molecular dynamics (MD) simulation as an effective numerical investigation approach has also been frequently utilized.[19-22] For example, through the study of the vibration of the Si NW by MD simulation, Cao et al.[23] found that the low quality factor is resulted due to the energy loss coming from atomic friction and nonlinearity. Using similar approach, Kim and Park[24] found that, the tensile strain could effectively mitigate both the intrinsic surface and thermal losses, and an improvement in Q-factor by a factor of 3~10 across a range of operating temperatures are observed for Cu NWs.

Some experimental and numerical works have also attributed to the study of the resonant frequency of NWs. For instance, by measuring resonance frequencies of the first and second flexural modes of silicon nitride cantilevers with different thickness, Gavan et al.[25] found that the determined effective Young's modulus strongly decreases for thickness below 150 nm, and the surface stress model is found to be in contradiction with experimental finds on the second flexural mode. Recently, the resonant property of strained single crystal Au NWs are studied by Olsson[26] through MD simulation. The results reveal that, eigenfrequencies of the fundamental mode follows the behaviour predicted by Euler-Bernoulli beam theory while those of high order modes are low in comparison. This phenomenon is considered as a result of the increasing influences from the shearing and rotary inertia for higher order resonant modes.[27]

We note that, although several numerical investigations of NWs under vibration have been reported. A comprehensive study of their vibration performance regarding different parameters, such as different initial excitations, different excitation magnitudes, and others is lack. For instance, in the work by Olsson et al.,[27] the higher vibration mode is excited by imposing distributed force to the atoms in a selected region. Due to the fact that improper force magnitude might cause the atomic interactions broken before the simulation begins, which makes this process as a trial and error procedure. Therefore, a fundamental numerical and theoretical study of the vibrational properties of Ag NWs is presented in this work. We begin from the discussion of two kinds of excitations, i.e., velocity excitation and displacement excitation. The influences from excitation magnitudes, temperatures are investigated. Specifically, the beat phenomenon is observed driven by a relatively large single excitation and its triggering mechanisms are discussed. In addition, a new loading schematic is proposed, which could excite the NW under a multi-vibration vibration mode and arrows the coexistence of all following low vibration modes.

## II. NUMERICAL AND THEORETICAL FUNDAMENTS
### A. Molecular Dynamics Implementation

The vibration tests of Ag NWs with end conditions as clamped-clamped (C-C) and cantilevered or clamped-free (C-F) are carried out by MD simulations. The Large-scale Atomic/Molecular Massively Parallel simulator (LAMMPS)[28] is employed to perform the MD simulation. Square cross-section Ag NW with the initial atomic configuration positioned at perfect FCC lattice site is considered, and the $x$, $y$ and $z$ coordinate axes represent the lattice direction of [100], [010] and [001], respectively. The simulation model is shown in Fig. 1, with $h$ denotes the NW cross-sectional size, and $L$ denotes the length. The Ag lattice constant $a$ is chosen as 0.409 nm.[29] During simulation, the NW is divided into two regions, including boundary region and mobile region. The C-C boundary condition (BC) is realized by fixing NW's two ends in all three dimensions ($x$, $y$ and $z$). For the C-F BC, one end is fixed in all three dimensions with the other free. No periodic boundary condition is adopted.

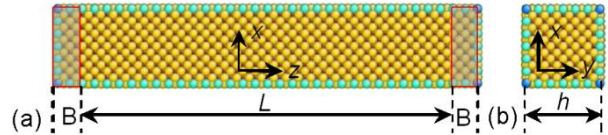

**Fig. 1.** Schematic of a clamped-clamped Ag NW model. Boundary regions 'B' are fixed in all directions, with the rest as the deformation region. The NW has a square cross-section with the lateral size of $h$, and the NW length is denoted as $L$. (a) Front sight view. (b) Cross-section view.

The embedded-atom-method (EAM) potential developed by Foiles et al.[30] is used to describe the Ag atomic interactions in these simulations, which was fitted to a group of parameters, including cohesive energy, equilibrium lattice constant, bulk modulus, and others.[31]

$$E_{tot} = \sum_i F(\rho_i) + \frac{1}{2}\sum_i \sum_j V(r_{ij}) \ , \rho_i = \sum_j \Phi(r_{ij}) \tag{1}$$

here $V$, $F$, $\rho$ are the pair potential, the embedded energy, and the electron cloud density, respectively. $i$ and $j$ are the number of atoms, and $r_{ij}$ is the distance between them. The equations of motion are integrated with time using a Velocity Verlet algorithm.[28] In order to recognize the plastic deformation that might caused by excessive initial excitation, the centro-symmetry parameter (*csp*) is utilized, which is defined by[32]

$$csp = \sum_{i=1,6} |\mathbf{R}_i + \mathbf{R}_{i+6}|^2 \qquad (2)$$

where $\mathbf{R}_i$ and $\mathbf{R}_{i+6}$ are vectors corresponding to the six pairs of opposite nearest neighbors in FCC lattice. The *csp* value increases from zero for perfect FCC lattice to positive values for defects and for atoms close to free surfaces.

## B. Theoretical Basics

As aforementioned, NW is usually operated as a resonating beam in NEMS, thus the sharpness of the resonant peak is crucial for high resolution or high accuracy of measurement. The parameter to indicate the sharpness of a resonance curve is the quality (Q)-factor of the system. There are currently several definitions for the Q-factor, which are equivalent for slight damping. Typically, Q-factor is defined as the ratio between the total system energy and the average energy loss in one radian at resonant frequency,[33] that is

$$Q = 2\pi \frac{E}{\Delta E} \qquad (3)$$

where $E$ is the total energy of the vibration system and $\Delta E$ is the energy dissipated by damping in one cycle of vibration. Assume Q-factor is constant during vibration, then after $n$ vibration cycles, the maximum energy $E_n$ is related to the initial maximum energy $E_0$ by[34]

$$E_n = E_0 (1 - 2\pi / Q)^n \qquad (4)$$

According to the Euler-Bernoulli beam theory, the governing partial differential equation for the beam under free vibration is[35]

$$EI \frac{\partial^4 w}{\partial z^4} + \rho A \frac{\partial^2 w}{\partial t^2} = 0 \qquad (5)$$

where $EI$ is the flexural rigidity, $\rho$ is density and $A$ is the cross-sectional area of the NW. $w(z,t)$ is the NW transverse displacement. By using the corresponding BCs, a uniform expression of the vibration frequency can be determined for both C-C and C-F NWs by

$$f_n = \frac{\omega_n}{2\pi L^2} \sqrt{\frac{EI}{\rho A}} \qquad (6)$$

where $\omega_n$ is the eigenvalue obtained from the characteristic equations of $\cos(\sqrt{\omega_n})\cosh(\sqrt{\omega_n}) = 1$ and $\cos(\sqrt{\omega_n})\cosh(\sqrt{\omega_n}) = -1$ for C-C and C-F NWs, respectively.

Following studies will focus on the discussions on the Q-factor and natural (resonance) frequency of the Ag NW. Specifically, due to the energy-preserving (NVE) ensemble adopted during NW vibration, the lost of potential energy (*PE*) must be converted to kinetic energy (*KE*). Thus, the time history of the external energy (*EE*) will be tracked for the calculation of Q-factor. *EE* is defined as the difference of *PE* before and after the initial excitation is applied to the NW.[24] To determine the vibration frequency from the simulation results, the Fast Fourier transform (FFT) is applied, which provides a fast computation of the discrete Fourier transform (DFT).[36] For the theoretical calculation of the natural frequency for the Ag NW, Young's modulus *E* is chosen as 76 GPa, and density $\rho$ equals $1.05 \times 10^4$ kg/m$^3$ according to Shenoy.[37] The area (*A*) and moment of inertia (*I*) is calculated by considering the NW as a continuum structure, i.e., $A=h^2$ and $I=h^4/12$ (*h* is the square cross-sectional size).

## III. RESULTS AND DISCUSSION

We begin this work by the discussion of two kinds of successfully implemented initial vibration excitations, i.e., velocity excitation and displacement excitation. To achieve this, a group of relatively small NWs with C-C BCs are considered, and their sizes are chosen as 6*a*×6*a*×34*a*. After that, another two groups of NWs are adopted to investigate the thermal effect to the natural frequency of the Ag NW. Each group possesses six NWs, and their sizes are chosen as 6*a*×6*a*×124*a* and 6*a*×6*a*×74*a*, respectively. The former group is set with C-C BCs and the latter one with C-F BCs. To note that, every clamped boundary region contains two lattice constants length that being set rigid, which makes the effective length of the NW equal to the initial length excludes those boundary regions. For instance, the 6*a*×6*a*×124*a* Ag NW has an effective slenderness ratio (*L*/*h*) of 20. A comprehensive discussion of the higher vibration modes is presented in the end. NWs with the slenderness larger than 10 are selected under the assumption that these NWs could be simplified as thin beams.[38]

### A. Velocity Excitation

An initial velocity is firstly adopted to drive the vibration of NW. Clamped-clamped (C-C) Ag NW with the size of 6*a*×6*a*×34*a* is considered. Basically, NWs are first relaxed to a minimum energy state using the conjugate gradient energy minimization and then the Nose-Hoover thermostat[39, 40] is employed to equilibrate NWs at 0.1 K under the ensemble of NVT. After the system reached the thermal equilibrium, a sinusoidal transverse velocity field is applied to the NW. The vibration of the NW is then achieved under a constant energy (NVE) ensemble. This procedure is similar as applied by previous researchers.[24, 41]

For discussion convenience, the initial velocity excitation is expressed as $u'(z) = \lambda \sin(kz)$, where $\lambda$ is the amplitude of initial velocity, and *k* equals $\pi/L$. *u*(*z*) is the displacement from the straight NW position in *x* direction of the Ag NW, and *u'*(*z*) is the corresponding velocity. Fig. 2a presents the time history of *EE* during the vibration with the initial velocity amplitude equals 1 Angstrom/psec. The gradual changing of the *EE* peak value indicates that the nonlinear vibration has not been excited. Particularly, the Q-factor is estimated around $5.4337 \times 10^4$, which is comparable with the value reported by Kim and Park.[24] Fig. 2b shows the first half of the periodogram regarding the power of the DFT versus frequency. As is seen, only one frequency component is identified, which is about 70.31 GHz. Since this is the vibration frequency of the external energy, which means the actual first order natural frequency ($f_1$) of the NW is around 35.15 GHz.

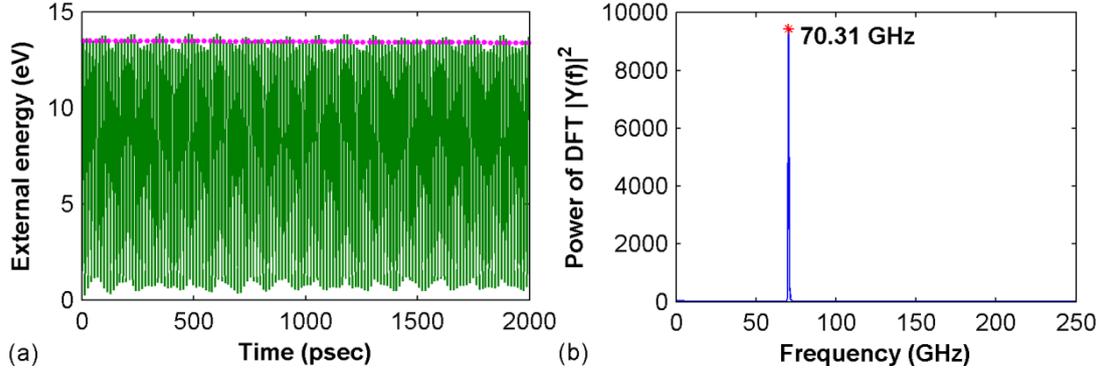

**Fig. 2.** Simulation results of the $6a \times 6a \times 34a$ Ag NW with the initial velocity amplitude equals 1 Angstrom/psec under the temperature of 0.1 K. (a) The time history of external energy (*EE*) during vibration. Circle markers highlighted the magnitudes of *EE* during each vibration circle. (b) First half of the periodogram, regarding the power of discrete Fourier transformation (DFT) versus frequency.

For the purpose of examining the influence from different amplitudes of the initial excitations, several different amplitudes $\lambda$ ranging from 0.05 Angstrom/psec to 2.5 Angstrom/psec are tested. Fig. 3a shows the natural frequency extracted using FFT from the simulation results. Evidently, NWs with the velocity amplitude values below 1.5 Angstrom/psec appears an identical $f_1$ as 35.15 GHz. When $\lambda$ reaches 1.75 Angstrom/psec, an enhanced frequency about 37.84 GHz is observed. Further increase of $\lambda$ finally leads the decrease of $f_1$. Atomic configurations reveal that for $\lambda$ larger than 1.75 Angstrom/psec, different amount of initial plastic deformation is introduced to the NW by the initial excitation. Fig. 3b presents the record of *EE* when $\lambda$ equals 2 Angstrom/psec, whose magnitudes reveals a nonlinear fashion against with the time. From Fig. 3c, an intrinsic stacking fault (SF) is observed, which appears intact during the whole vibration process. It is observed that, the external energy received an obvious attenuation during the generation of this stable SF (at the beginning of vibration in Fig. 3b). As revealed in Fig. 3a, the frequency of the NW when $\lambda$ equals 2 Angstrom/psec is even larger than the NW without plastic deformation, indicating plastic deformation sometimes might increase the NW's natural frequency.

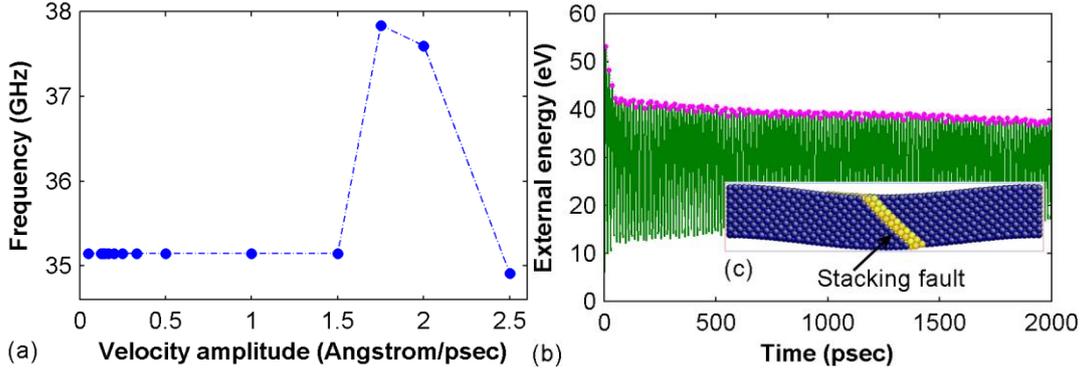

**Fig. 3.** Simulation results of the $6a \times 6a \times 34a$ Ag NW under velocity excitation at the temperature of 0.1 K. (a) The first vibration frequency as a function of the velocity amplitude. (b) The time history of external energy (*EE*) during vibration for the Ag NW when the initial velocity amplitude equals 2 Angstrom/psec. Circle markers highlighted the magnitudes of *EE* during each vibration circle. The inset figure (c) represents the atomic configurations of this NW at 100 psec, atoms with the *csp* value between 0.5 and 12 are visualized.

### B. Displacement Excitation

The displacement excitation is also successfully implemented to drive the vibration of NW. Clamped-clamped (C-C) Ag NW with the size of $6a \times 6a \times 34a$ is considered. In detail, after the NW is equilibrated at 0.1 K under the ensemble of NVT, a virtual cylindrical tip is employed to bend the NW to a pre-set deflection. To note that, this virtual tip has no real shape, which is similar as the spherical indenter tip applied during nanoindentation studies[42, 43] and the cylindrical tip applied during bending studies.[44] Basically, the virtual tip will exert a repulsive force between the tip and the NW, which is expressed by $F(r) = -c(r-R)^2$, here $c$ is the specified force constant, $r$ is the distance from the atom to the centre of the tip and $R$ is radius of the tip. When the deflection or displacement arrives at the required value, the virtual tip is removed, and thus, a sinusoidal transverse displacement field is applied to the NW. Specifically, the displacement stimulus could be written as $u(z) = \delta \sin(kz)$, where $\delta$ is the maximum deflection value, and $k$ equals $\pi/L$. Upon this initial displacement filed, the vibration of the NW is then achieved under a constant energy (NVE) ensemble.

Comparing with the vibration results driven by the velocity stimulus, a similar *EE* time history curve is tracked. Fig. 4a presents the *EE* versus time curve when $\delta$ equals 5.91 Angstrom. The *EE* magnitude is found to decrease linearly with the increase of time, and the Q-factor is calculated around $3.5638 \times 10^4$. Fig. 4b describes the frequency obtained from the vibration of Ag NWs under different initial deflection values. Like the results in Fig. 3a, the first order natural frequency $f_1$ does not change at the beginning, and it increases after the initial stimulus passes a certain value, the increasing initial displacement eventually causes obvious reduction to the $f_1$. In particular, the initial frequency values are uniformly estimated as 35.15 GHz for the considered Ag NWs under both velocity and displacement excitations. Though only several discrete amplitude values have been tested, we could still expect that the changing tendency, as well as the values of $f_1$ in Figs. 3a and 4b is coincident under these two kinds of excitations. Such consistent results in the other hand approve the accuracy of the current vibration simulation.

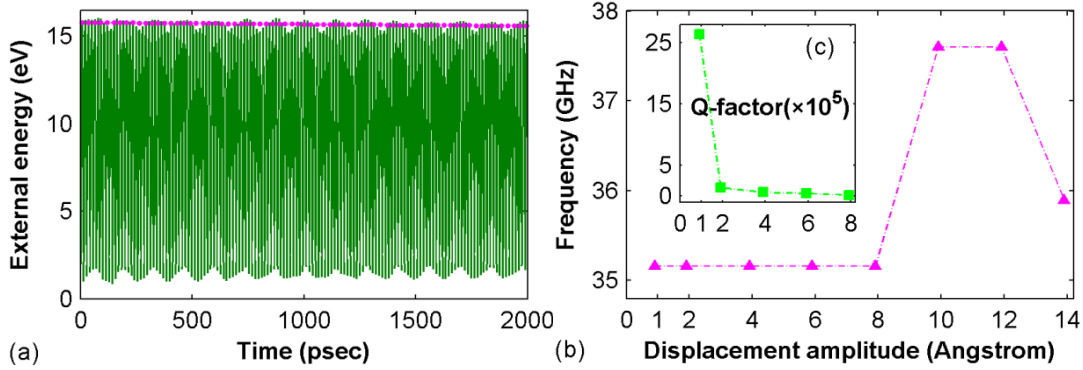

**Fig. 4.** Simulation results of the $6a \times 6a \times 34a$ C-C Ag NW under displacement excitation at the temperature of 0.1 K. (a) The time history of external energy (*EE*) during vibration for the Ag NW with the initial displacement amplitude equals 5.91 Angstrom. Circle markers highlighted the magnitudes of *EE* during each vibration circle. (b) The first vibration frequency as a function of the displacement amplitude. The inset figure (c) represents the Q-factor as a function of the displacement amplitude.

Fig. 4c depicts the changing trend of the Q-factor for Ag NWs under different displacement excitations. Since there are plastic deformation observed when $\delta$ is larger than 9.91 Angstrom, which leads to a nonlinear pattern of the *EE* magnitude, thus, only Q-factors with $\delta$ under this value are estimated. Overall, due to the increase of initial deflection, an evident degradation of Q-factor is observed. For instance, for $\delta$ equals 0.91 Angstrom, the Q-factor is estimated around $10^6$, whilst for $\delta$ equals 3.91 Angstrom, the Q-factor is only around $5.684 \times 10^4$. These observations are consistent with the simulation results under the velocity excitation. The degradation of Q-factor due to the increase of initial excitation amplitudes could be explained in two aspects. As revealed by the *EE* time history in Figs. 2a, 3b and 4a, larger initial excitations means larger energy input to the vibration system, which will lead the NW vibrating around a higher temperature (though their initial temperature are the same as 0.1 K). For example, the Ag NW is found to oscillate around 0.37 K and 13.5 K for $\delta$ equals 0.91 Angstrom and 5.91 Angstrom, respectively. Besides, larger initial excitation indicates severer deformation during vibration, which would require more affluent atomic interactions. Therefore, considering both the thermoelastic damping[45] and atomic friction[23] during vibration, it is reasonable to observe the NW driven by a larger excitation possesses a smaller Q-factor.

## C. Beat Vibration

One interesting phenomenon unveiled by the above vibration test is the beat vibration. Fig. 5a illustrates the *EE* time history when $\delta$ equals 13.91 Angstrom. Since the NW is approaching a stable deformed structure with the existence of SF at the beginning of vibration, the amplitude of *EE* decays greatly with time. After the initial attenuation, the external energy amplitude exhibits a periodic pulsation pattern. According to the vibration theory, this changing trend is usually generated by a combination of two simple harmonic vibrations with almost same frequency, namely the beat vibration. To probe our concern, the atomic configurations are inspected carefully. As demonstrated in Figs. 5b and c, the Ag NW is found to vibrate not only in the pre-set direction (*x* direction in Fig. 5b), but also in another lateral direction (*y* direction in Fig. 5c). This observation signifies that the Ag NW does vibrate in both lateral directions. Explanations are given as below.

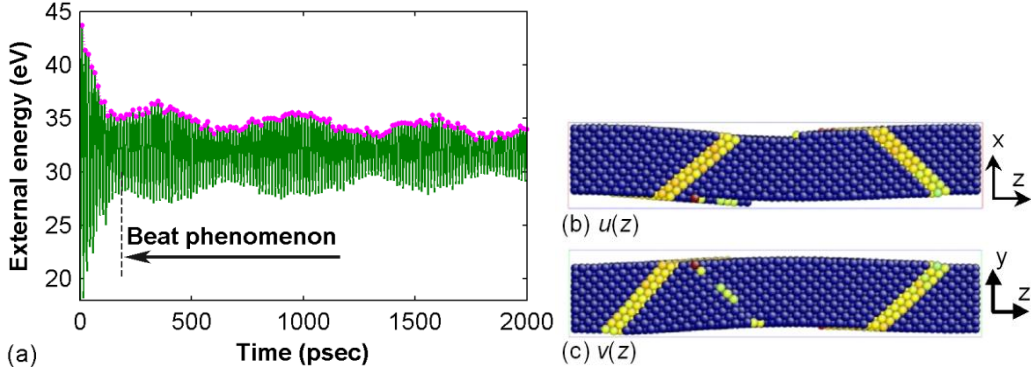

**Fig. 5.** Simulation results of the $6a \times 6a \times 34a$ C-C Ag NW under the temperature of 0.1 K with the initial displacement amplitude equals 13.91 Angstrom. (a) The time history of external energy (*EE*) during vibration. Circle markers highlighted the magnitudes of *EE* during each vibration circle. (b) Atomic configuration at 1600 ps reveals the vibration $u(z)$ in *x* direction. (c) Atomic configuration at 1600 psec reveals the vibration $v(z)$ in *y* direction. In the atomic configurations, atoms with the *csp* value between 0.5 and 12 are visualized.

It is certain that, for the perfect Ag NW free of plastic deformation, the *x* and *y* direction of the Ag should possess same flexural rigidity (because of its symmetric cross-section). However, due to the presence of SFs, the crystal structure of the cross-section is no longer symmetric, which results in the minor offset between the natural frequencies in these two directions. This argument is approved from the frequency spectrogram derived from the FFT analysis, which detected two close frequencies values as 35.89 GHz and 36.62 GHz. Although this fact satisfies the fundamental requirement for a beat vibration, it is still unclear how the single excitation drives two harmonic vibrations of the NW. The authors suggest that, the existence of the SFs induces the decomposing of the initial excitation, which generates two vibration components. Certainly, the beat vibration should not be limited by the displacement excitation. Its occurrence is found as well when the velocity amplitude exceeds 2.0 Angstrom/psec. In summary, to engender the beat vibration, the initial excitation should be large enough, that arrows sufficient plastic deformation to break the symmetry in the two lateral directions of the NW.

The beat phenomenon is further confirmed by considering a C-C Ag NW with a rectangular cross-section. The size of the NW is chosen as $6a \times 6.5a \times 34a$. After the NW arrived at the equilibrium state, two uniform initial velocity excitations are applied simultaneously in both *x* and *y* directions. Fig. 6a represents the *EE* time history during vibration. Just as expected, a typical beat vibration is reproduced. From Fig. 6a, the beat period (*Tb*) of the external energy is estimated around 230 psec. The emergence of the extremely small amplitude in Fig. 6a suggests that the frequencies of the two simple harmonic vibration components are very close to each other, which agrees with the FFT analysis in Fig. 6b. Two close frequencies as 70.8 GHz and 75.2 GHz are identified, which indicate a beat period around 227.3 psec for the external energy. It should be pointed out that, since the cross-section has changed to rectangle, thus, the moments of inertia in *x* and *y* direction are expressed by $I=bh^3/12$ and $I=hb^3/12$, respectively (*h* and *b* are the two lateral sizes in the *x* and *y* direction, respectively). Apparently, these two expressions indicate a larger moment of inertia than the previous square cross-section ($I=h^4/12$) when *b* is larger than *h*. Therefore, the slight increase of the frequency from 35.15 GHz (Figs. 3a and 4b) to 35.4 GHz and 37.6 GHz (Fig. 6b) is explainable. In conclusion, the beat vibration is successfully reproduced, which confirms the beat phenomenon observed in the previous case (driven by a single relatively large initial stimulus).

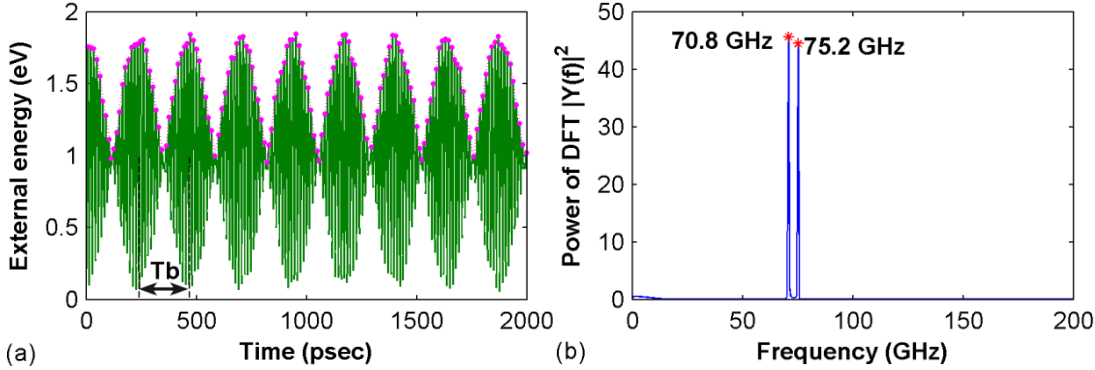

**Fig. 6.** Simulation results of the *6a×6a×34a* C-C Ag NW under two uniform velocity stimuli (velocity amplitudes of 0.75 Angstrom/psec) in both *x* and *y* directions at the temperature of 0.1 K. (a) The time history of external energy (*EE*) during vibration. Circle markers highlighted the magnitudes of *EE* during each vibration circle. (b) First half of the periodogram regarding the power of DFT versus frequency.

One more point is about the Q-factor estimation. As seen in both Figs. 5a and 6a, the *EE* magnitudes are highly nonlinear. Assume that the external energy of the two simple harmonic vibrations could be described by $Ae^{nt}\sin(2\pi\omega_a t)$ and $Ae^{nt}\sin(2\pi\omega_b t)$, respectively. Here, $\omega_a$ and $\omega_b$ are two close frequencies. Same amplitude $Ae^{nt}$ and zero initial phases are assumed, where *n* relates with the system damping coefficients, and $e^{nt}$ represents the amplitude decays exponentially. Combine this two vibration components leads:

$$EE = 2Ae^{nt}\cos(2\pi\frac{\omega_a-\omega_b}{2}t)\sin(2\pi\frac{\omega_a+\omega_b}{2}t) \tag{7}$$

Therefore, the overall amplitude equals:

$$A_o = 2Ae^{nt}\cos(2\pi\frac{\omega_a-\omega_b}{2}t) \tag{8}$$

Equation 8 suggests that the overall amplitude not only decays exponentially but also oscillate under the beat frequency of ($\omega_a-\omega_b$). However, for each vibration component, the amplitude still decays exponentially, which indicates a constant Q-factor. Therefore, the conclusion is achieved that the nonlinearity of the *EE* magnitude does not always mean a non-constant Q-factor. For a vibration contains two or more vibration components, the Q-factor should be calculated according to other definitions, e.g., the mathematical definition, which is directly related to the damping ratio of the vibration system. In addition, from the results in Sec. III A & B, the Q-factor also appears vulnerable to the excitation magnitudes.

## D. Temperature Influence

Further interest is laid on the consistency between simulation results and predictions from the classical Euler-Bernoulli beam theory (Eq. 6). Although there are already certain works studied the comparison between the MD results and theory calculations, the comprehensive investigation of the temperature influence is still untouched. In this section, the vibrations of both C-C and C-F Ag NWs under various temperatures, ranging from 0.1 K to 400 K are studied. Specifically, the Ag NW sizes are selected as $6a \times 6a \times 124a$ and $6a \times 6a \times 74a$ for the C-C and C-F BCs, respectively. The simulation setting is similar as introduced in Sec. III A. Since the thermal effect on the Q-factor is already well presented by previous researchers,[24, 34] hence, the following discussions will emphasize on the natural frequency of the NW.

Figure 7a reveals the first order natural frequency of the Ag NW as a function of temperature. According to Eq. 6, the first order natural frequency $f_1$ is estimated to be 2.82 GHz for the considered C-C NW (where $\omega_1 = 22.37$). As shown in Fig. 7a, $f_1$ decreases with the increase of temperature, indicating Young's modulus of the NW is decreasing against with the temperature. Such conclusion is in accordance with previous results from NWs under tensile deformation.[43, 46] Furthermore, the frequency values that extracted from simulation results appear larger than the theoretical prediction when the temperature is lower than 150 K. For instance, for the temperature of 0.1 K, $f_1$ is around 3.05 GHz, which is about 8.16% larger than the theoretical value. On the contrary, $f_1$ is about 26.24% smaller than the theoretical value at the temperature of 400 K, which is only about 2.08 GHz.

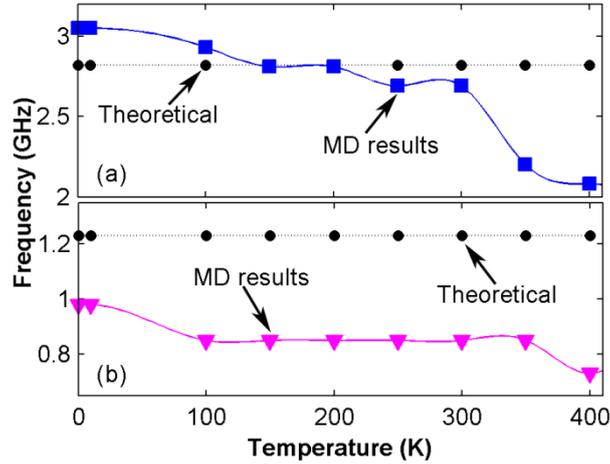

**Fig. 7.** Simulation results of thin Ag NWs under velocity stimuli at various temperatures ranging from 0.1 K to 400 K. (a) The first order natural frequency as a function of temperature for the C-C NWs. (b) The first order natural frequency as a function of temperature for the C-F NWs.

For the C-F NW, the first order natural frequency is calculated as 1.23 GHz from Eq. 6 (where $\omega_1 = 3.514$). As illustrated in Fig. 7b, comparing with the results from the C-C NWs, a more gradual decreasing of $f_1$ is accompanied with the increase of temperature. Moreover, all the frequency values deduced from the testing results are smaller than the theoretical calculation. For example, for the lowest examined temperature (0.1 K), the frequency (about 0.98 GHz) is found to be 20.33% smaller than the theoretical value. To mention that, the deviations of the frequency for the C-C and C-F NWs between the numerical values and theoretical predictions in Fig. 7 are agreed with the recent theoretical work by He and Lilley.[16] They suggested that, a positive surface stress decreases the resonance frequencies of the cantilever (C-F) NWs and increases the resonance frequencies of clamped-clamped (C-C) NWs. In all, the first order natural frequency is found to decrease with the increase of temperature for NWs with both C-C and C-F BCs, and its value is comparable with the beam theory prediction.

### E. High Vibration Modes

We close this work by the discussion of high vibration modes. The C-C Ag NW with the size of $6a \times 6a \times 124a$ is considered. In Sec. III A and B, we find that increase the amplitude of the sinusoidal velocity or displacement excitations does not excite high vibration mode. Therefore, according to the profile of the beam under high vibration mode, the velocity excitations include $u'_2 = \lambda_2 \sin(2kz)$, $u'_3 = \lambda_3 \sin(3kz)$ and $u'_4 = \lambda_4 \sin(4kz)$ have been employed, which have successfully excited the second, third and fourth vibration modes, respectively. In consideration of the possible influence from different excitation magnitudes, we first examine the thin NWs under the first vibration mode driven by several different velocity excitations. Fig. 8a shows the first order natural frequency of the Ag NW as a function of excitation amplitude. Different from the results from short NWs ($6a \times 6a \times 34a$) in Figs. 3a and 4b, a parabolic relationship is found between the frequency and the velocity amplitudes. Theoretically, larger excitation amplitude means larger deflection during vibration (before the plastic deformation happens). Aware that dynamic elongation of the NW is inherently induced during the vibration of NW, as a result, a nonlinear term that stemmed from this elongation can be incorporated to the governing equation (Eq. 5) as

$$EI \frac{\partial^4 w}{\partial z^4} - T \frac{\partial^2 w}{\partial x^2} + \rho A \frac{\partial^2 w}{\partial t^2} = 0 \qquad (9)$$

where $T$ is the dynamic force induced by the dynamic stretching of the NW, it is approximated as[47]

$$T = EA\varepsilon_z = \frac{EA}{L} \int_0^{L/2} (\frac{dw}{dz})^2 dz \qquad (10)$$

Solving Eq. 9 could be a tedious work, however, according the early work by Bouwstra,[48] this dynamic axial force would increase the natural frequencies on the order of $(w_{max}/h)^2$. Therefore, it is reasonable to observe an increasing frequency along with the increase of excitation magnitudes as revealed in Fig. 8a. Furthermore, according to our recent work regarding the bending study of Ag NWs,[44] the short NWs (with low slenderness ratio) yields before the nonlinear effect becomes crucial, whilst, the long NWs exhibit a cubic like force versus displacement curve before yielding. This phenomenon suggest that the dynamic axial force exert less or insignificant influence to short NWs, which explains the uniform frequency values determined for the short NWs considered in Sec. III A and B. In addition, the thin NWs with the existence of moderate plastic deformation (e.g., stacking faults) are also found to possess larger first order natural frequency than some other NWs without plastic deformation, which suggests the same conclusion that, moderate plastic deformation would increase the first natural frequency.

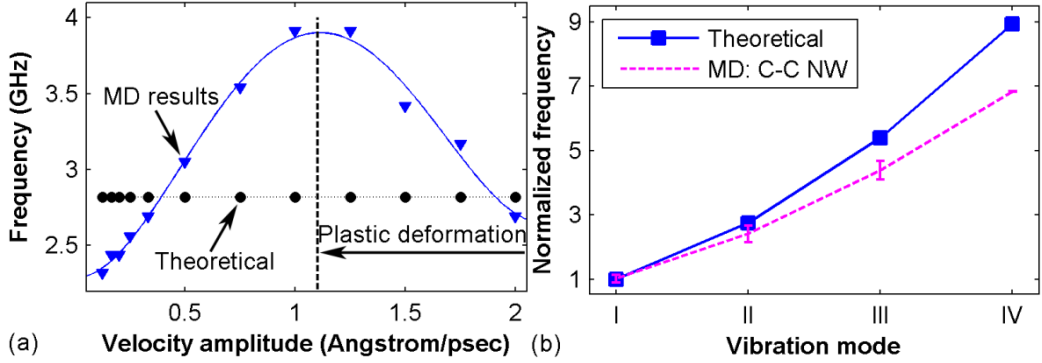

**Fig. 8.** Simulation results of C-C $6a \times 6a \times 124a$ Ag NWs under velocity stimuli at the temperature of 10 K. (a) The first vibration frequency as a function of the velocity amplitude. (b) Comparison of the normalized frequency for different vibration modes between simulation results and theoretical calculations. Error bars describe the standard deviation of the normalized frequency under different velocity magnitudes.

Figure 8b shows the normalized frequency obtained from the MD simulations for the first (I), second (II), third (III) and forth (IV) vibration modes. The normalized frequency is defined as $f_N = f / f_{o1}$, here $f_{o1}$ is the theoretical first order natural frequency (2.82 GHz for the C-C $6a \times 6a \times 124a$ Ag NW). For each high vibration mode (II, III and IV), six cases (with a range of velocity excitation magnitudes) have been studied to conquer the potential influence from the excitation magnitude. The standard deviation of the frequency values are calculated, as described by the error bars in Fig. 8b. Basically, except the first vibration mode, all higher vibration modes performed by the simulation tend to a lower frequency value than the theoretical calculations. Moreover, the shift between the theoretical prediction and the simulation results rises with higher vibration mode. These observation are agreed with the results reported by Olsson,[26] i.e., the higher order modes tend to be low in comparison to Euler-Bernoulli beam theory predictions. Such phenomenon is supposed resulting from the naturally omitted contributions from the shear deformation and rotational inertia, which are expected to exert considerable influence to high vibration modes.[27, 48]

In the end, we introduce a new loading schematic to excite high vibration modes. As mentioned by previous researchers,[27] when the high vibration modes are excited, only a single or pure mode can be found. This argument also agrees with our above vibration tests, e.g., $u'_2$ could only excite the second vibration mode, and no first vibration component is resolved by the FFT analysis. The phenomenon contradicts with the vibration theory, since high vibration modes require more excitation energy, which means high vibration mode is always accompanied by the following low vibration mode. To overcome this problem, a series of vibration testing are conducted. We find that a proper combined initial stimulus could easily drive the NW under a multi-modes vibration. For instance, applying the initial excitation $u'_{12} = \lambda_1 \sin(kz) + \lambda_2 \sin(2kz)$ to the Ag NW, both first and second vibration modes can be excited. Fig. 9a shows the frequency spectrogram of a C-C Ag NW under a multi-stimulus $u'_{123}$ (which contains three fundamental excitations). Three peak values in Fig. 9a correspond to the first three natural frequencies, which are coincident with the values under pure mode vibration.

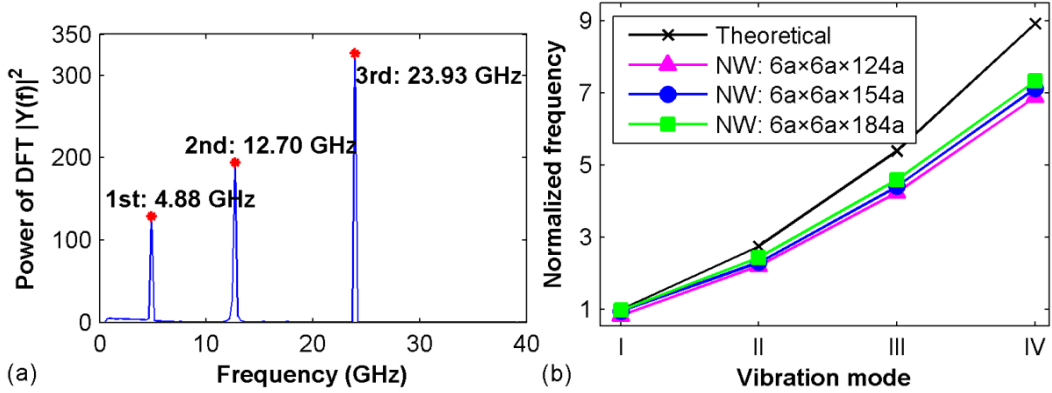

**Fig. 9.** Simulation results of thin Ag NWs under a combined velocity excitation at the temperature of 10 K. (b) First half of the periodogram, regarding the power of discrete Fourier transformation (DFT) versus frequency for the $6a \times 6a \times 124a$ Ag NW. (b) Comparison of the normalized frequency at different vibration modes between simulation results and theoretical calculations for NW with three different lengths, i.e., 124$a$, 154$a$ and 184$a$.

Several vibration tests are conducted to verify such loading schematic. NWs with the length of 124a, 154$a$, and 184$a$ are considered. A multi-stimulus which contains four fundamental excitations is adopted. Fig. 9b compares the normalized frequency extracted from the testing results with the theoretical predictions. In general, four frequency components are identified for each NW, and their values agree well with the theoretical prediction. The phenomenon that, the higher order vibration modes tend to be low in comparison to Euler-Bernoulli beams theory predictions is also observed. In addition, the frequencies are observed closer to the theoretical calculations for longer NWs, which suggest that the shear deformation and rotational inertia exert less influence to longer NWs. To note that, during all the discussions of the theoretical calculation in this work, no assumptions about the surface effect have been made. Considering the good agreement of the current MD results with the classical beam theory, we expect that the augmented beam theory with the contribution of surface effect would attain better results, especially for the high vibration mode.

## IV. CONCLUSION

The vibration of Ag nanowires (NWs) is well reproduced by molecular dynamics (MD) simulation. A comprehensive analysis of the quality (Q)-factor, natural frequency, beat vibration, as well as high vibration mode has been conducted. Considered NWs include short and long Ag NWs, and two kinds of boundary conditions (i.e., clamped-clamped and clamped-free) are adopted. The Ag NW's vibration performances under different excitation magnitudes and different temperatures are studied. Comparisons of the natural frequencies for different vibration modes between the simulation results and theoretical predictions by the classical Euler-Bernoulli beam theory are carried out. Major conclusions are summarized as below:

(1). The velocity excitation and displacement excitation have been successfully implemented to achieve the vibration of NW, which yield consistent results, i.e., the increase of the initial excitation amplitude will lead to a decrease to the Q-factor, and moderate plastic deformation could increase the first order natural frequency;

(2). The beat vibration of the NW driven by a single relative large excitation or two uniform excitations in both two lateral directions is observed. It is suggested that the nonlinear changing trend of the external energy magnitudes does not necessary mean a non-constant Q-factor;

(3) The first order natural frequency is observed to decrease against with the increase of temperature, indicating Young's modulus decreases with the increase of temperature. Furthermore, comparing with the predictions by Euler-Bernoulli beam theory, the MD simulation provides a larger and smaller first vibration frequency for the double clamped and clamped-free thin Ag NWs, respectively.

(4). For thin NWs, the first order natural frequency exhibits a parabolic relationship with the excitation magnitudes. It is supposed that, in the elastic regime, larger excitation would induce larger vibration amplitude, meaning larger dynamic elongation (induced inherently by the vibration). As a result, the influence to the natural frequency from such dynamic elongation increases.

(5) The frequencies of the higher vibration modes tend to be low in comparison to Euler-Bernoulli beam theory predictions, which are considered as a result of the increasing influence from the shearing deformation and rotational inertia.

(6). A combined initial excitation schematic is proposed, which could drive the NW under a multi-modes vibration and arrows the coexistence of all the following low vibration modes. For example, when the NW is under the third vibration mode, the first, and second vibration modes are also existed.

This work sheds lights on the better understanding of the mechanical properties of NWs, and the simulation techniques and analysis methods delineated in this work should also be applicable to other NWs (such as pinned-pinned thin NWs, as well as thick NWs with smaller slenderness). The study in this work would also benefit the increasing utilities of NWs in diverse nano-electronic devices.